\documentclass[aps,prl,twocolumn,nofootinbib, showpacs]{revtex4}
\usepackage{graphicx}
\usepackage{amsmath}
\bibpunct[, ]{[}{]}{,}{a}{,}{,}
\newcommand{\be}{\begin{equation}}
\newcommand{\ee}{\end{equation}}
\newcommand{\ab}{{\mathcal A}}  
\newcommand{\vS}{{\vec{S}}}
\begin{document}
\title{Dynamics of Competitive Evolution on a Smooth Landscape}

\author{Weiqun Peng}
\author{Ulrich Gerland}
\author{Terence Hwa}
\author{Herbert Levine}
\affiliation{Center for Theoretical Biological Physics and Department
of Physics, University of California at San Diego, La Jolla, CA
92093-0319}
\date{\today}
\begin{abstract}

We study competitive DNA sequence evolution directed by {\it in vitro}
protein binding.  The steady-state dynamics of this process is well
described by a shape-preserving pulse which decelerates and eventually
reaches equilibrium. We explain this dynamical behavior within a
continuum mean-field framework.  Analytical results obtained on the
motion of the pulse agree with simulations. Furthermore, finite
population correction to the mean-field results are found to be
insignificant. 

\end{abstract}

\maketitle Competitive evolution such as breeding has been practiced
for ages. With recent advances in molecular biology, this method is
widely used to develop novel proteins and DNA sequences for a
variety of applications~\cite{REF:Arnold2001}. The basic idea of
competitive molecular evolution is  straightforward: in each
generation, a number of molecules with certain desired
characteristics are selected from the population; they are then
diversified (via point mutation and/or
recombination~\cite{REF:Stemmer1994}) and amplified back to the
original population size. The ``speed'' of evolution as well as the
final equilibrium distribution depend on a variety of factors such as
the mutation rate, selection strength, molecule length, and
population size. A systematic quantitative understanding of these
dependencies is lacking thus far. Such understanding is not only of
theoretical interest, but also helpful in improving the efficiency of
the breeding processes. In this study, we develop a
theoretical model for the simplest type of competitive evolution
involving only point mutations on a smooth landscape.
We achieve an understanding of this model with concepts and techniques
developed in the study of front propagation~\cite{REF:Collet1990}.

To make the discussion concrete, we
focus on the \textit{in vitro} evolution of DNA sequences due to
competitive binding to proteins. An example of such a system is
the recent experiment of
Dubertret~\textit{et.\ al.}~\cite{REF:Dubertret2001},
where DNA sequences are selected competitively according to their
relative affinities for the \textit{lac}-repressor protein.
In this experiment, selection is accomplished
by coating a beaker with \textit{lac}-repressor molecules
followed by subsequent washing, so that only the
strongly-bound sequences remain. Mutation and amplification
are then accomplished by multiple stages of polymerase
chain reaction~\cite{REF:Lodish2000}.
While the experiment of Ref.~\cite{REF:Dubertret2001} easily
accomplished the goal of finding the best binding sequence
starting from a pool of random sequences in a few generations,
the shortness of the binding sequence [$20$ base pairs (bp)]
makes it difficult to explore the interesting
dynamics of the competitive evolution process. In our study
we consider the evolution process of
Ref.~\cite{REF:Dubertret2001} applied to much longer sequences
so that that the steady state dynamics can be examined.
An example of such a system might be the histone-octamer, which
is known to  bind DNA sequences of $147$\;bp~\cite{REF:Widom1999}.

We consider a pool of $N$ DNA sequences of length $L$. Each sequence
 $\vS=(b_1,b_2, ...,b_L)$ of nucleotides $b_i$ is subject to
independent single-nucleotide
mutations at a rate $\nu_0\ll 1$ per nucleotide per
generation.  Selection is accomplished through protein-DNA binding.
Let the binding energy of a sequence $\vS$ to the protein be $E_{\vS}$
and let the fraction of such sequences in the pool be $n_{\vS}$.
Assuming thermodynamic equilibrium for the binding process, the
selection function is simply the binding probability, given by the
Fermi function~\cite{REF:VonHippel}
$P(E_{\vS}, \mu) = 1/[1 + \exp (E_{\vS} - \mu)]$,
where 
$mu$ is the chemical
potential and all energies are expressed in units of $k_{\rm B} T$.
Here $\mu$ serves as a (soft) selection threshold.
and is determined by the fraction $\phi$ of DNA
sequences that remain bound to the proteins after selection,
 i.e., $\sum_{\vS} \,P(E_{\vS},\mu) \,n_{\vS} = \phi$. It can
be controlled by either the number of available proteins or, as in
the experiment~\cite{REF:Dubertret2001}, by the
washing strength. The fraction $\phi$ can be varied from
$\phi\lesssim 1$ (weak competition)
to $\phi \gtrsim 0$ (strong competition). We define the evolution process
iteratively whereby in each round, $N$ daughter sequences are chosen
from the existing pool according to $P(E_{\vS},\mu)$, and then
point mutation are introduced with  rate $\nu_0$ to generate
the new sequence pool.

Finally we need to specify the
binding  energy $E_{\vS}$. We assume that each nucleotide taking
part in the binding contributes {\em independently}, and adopt a
\lq\lq two-state\rq\rq\ model~\cite{REF:VonHippel} which assigns an energy 
penalty $\epsilon$ (of the order of a few $k_{\rm B} T$'s)
for each nucleotide which does not match the one
the protein prefers. This form of binding energy has been shown to work
reasonably well for specific systems~\cite{fields} and has been argued
to hold for a wide class of regulatory proteins~\cite{gmh,nonspecific}. 
Given this energy model, a DNA
sequence with $r$ mismatches has a binding probability
$P(r, r_{0}) = 1\Big/\left[ 1 + e^ { \epsilon (r - r_{0})}
\right]$,
where $r_0 \equiv \mu/\epsilon$ is the selection threshold
in the ``mismatch space'' $r$.~\cite{REF:Gerland2001}.

\begin{figure}
\includegraphics[height=1.5in]{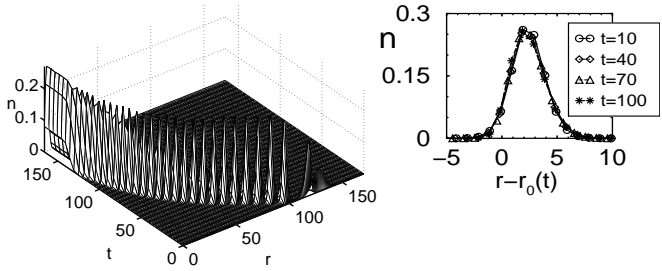}
\caption{\protect\flushing \label{shapeevolution} The
space-time trajectory of the mismatch distribution $n(r,t)$
according to the competitive evolution model with $\phi=0.1$.
The inset shows the distribution $n(r,t)$ at
generations $t=10, 40, 70, 100$, after the initial transient
period and before the distribution reaches equilibrium at $r\approx 0$.
These distributions  overlap upon shift by their
respective threshold $r_0(t)$, indicating the shape-invariance
of the pulse.}
\end{figure}

The above evolution model is easily implemented on a computer. We fix
three of the model parameters at $N=5\times10^5$, $L=170$ and
$\nu_0=0.01$ from here on, and  vary only the selection strength
through the choice of the selection stringency $\phi$. A typical
simulation result for strong selection ($\phi=0.1$) is shown as the
space-time plot of the mismatch distribution 
$n(r,t)\equiv \sum_{\vS} \,n_{\vS}\,\delta(E_{\vS}-r \epsilon)$ in
Fig.~\ref{shapeevolution}. We see that the distribution quickly forms
a  shape-preserving pulse (see the inset of
Fig.~\ref{shapeevolution}), which moves, decelerates, and eventually
reaches equilibrium in the neighborhood of the optimal sequence (at
$r=0$). Basically, the selection eliminates weak binders in the
population to improve the average binding energy, hence the selection
threshold $r_0$ is decreased in the next round, while the change of
$r_0$ further selects sequences with better binding energies.
Along with new variety generated by mutation,
 a {\it propagating pulse} results.

\setlength{\unitlength}{1mm}

We next investigate the dynamical behavior of the above evolution model
analytically using a mean-field description. It will be convenient to 
describe the dynamics in the mismatch space $r$. Let us first
consider the contribution from point mutation. For a
sequence of length $L$, ``alphabet size'' $\ab$ ($\ab =4$ for nucleotides)  
and $r$ mismatches, 
there are $L\cdot(\ab-1)$ ways to mutate to a new sequence 
via a single point mutation. Among them, there are
$(L-r)(\ab-1)$ ways to increase $r$ by one,
and $ r$ ways to reduce $r$ by one. Hence, a standard master equation can 
be written to describe the mutational dynamics of the distribution $n(r,t)$
in the mean field limit $N \gg 1$ \cite{REF:Peliti1997,REF:Drossel2001, REF:Gerland2001}.
The effect of selection/amplification process can be
phenomenologically modeled by an additive term proportional to
$\phi^{-1} P(r,r_0)$ for weak selection.  Further taking the continuum
limit in $r$ (valid in the limit of large $L$ and smooth population
distribution), we arrive at the following mean-field description for
$n(r,t)$~\cite{REF:Gerland2001}:
\begin{eqnarray} 
& &\partial_t n = \partial_r
\left[  \partial_r (D(r)\, n) - v(r) n \right] + \, U[r;n]\cdot n(r,t)
\label{EQ:MeanField1}\\
& &U[r;n] = \left[\phi^{-1} P(r,r_0(t)) -1\right]/\tau,
\label{EQ:U}
\end{eqnarray}
The first two terms on the right-hand side of (\ref{EQ:MeanField1})
result from the (conservative) mutational processes, with 
\begin{equation}
D(r) = \frac {\nu}{2} \left(1 - \frac{\ab - 2}{\ab - 1}\frac{r}{L}\right),
\quad
v(r) ={\nu} \left(1 - \frac{\ab}{\ab - 1}\frac{r}{L}\right)
\label{EQ:Dandv}
\end{equation}
being the ``diffusion coefficient'' and ``drift
velocity'' respectively~\cite{REF:Gerland2001}, $\nu\equiv \nu_0 L$. 
The $r$-dependences of $v$ and $D$  reflect the
different phase space volume for the different mismatches.
For example, the form of $v(r)$ ensures that the
the distribution approaches the maximum entropy point
with $\overline{r} = \frac{\ab-1}{\ab} L$ mismatches
by mutation alone.
The third term in Eq.~(\ref{EQ:MeanField1})
represents the effect of the selection/amplification, 
controlled by the growth function $U$ defined in
Eq.~(\ref{EQ:U}). (The factor $\tau \sim O(1)$ denotes generation time.)
 Competition is explicitly manifested in the $n$
dependence of the growth function $U$, via
the threshold $r_0(t)$ which is
determined from the condition $\phi = \int dr P(r,r_0(t)) n(r,t)$.
In Eq.~(\ref{EQ:U}),
an overall shift in $U$ by the constant $-1$ has been included to
 ensure that
the population size $N$ is {\em conserved} after selection/amplification, 
in accordance with the evolution process.
This shift produces the desired competitive effect that
individuals which bind better than the
threshold $r_0(t)$ are reproduced and those not meeting the
threshold decay away. In the actual analysis, we will
approximate the Fermi function $P(r,r_0)$ by a
step function $\Theta(r_0-r)$, which turns out not to 
affect the qualitative behavior.

We will see that the simplicity of the continuum mean-field equation
provides an analytic understanding of generic features of the
evolutionary dynamics, including the existence of the decelerating,
shape-preserving population pulse; it also provides an analytical
estimate of the smallness of the finite-$N$ correction.  However that
quantitative differences do exist between our simplified description
and the breeding schemes employed in the simulations and experiments,
due to the phenomenological nature of simplified description, and 
the continuum approximations used in both the mismatch space
and time.  (A quantitatively more accurate approach has
recently been developed by Kloster and Tang~\cite{REF:Kloster2}.)


We start with the simplest case of infinite sequence length (while
keeping $\nu$ a finite constant), yielding constant coefficients
$D(r)= D$ and $v = \nu$.  Making the Ansatz in Eq.~(\ref{EQ:MeanField1}) that
the distribution $n(r,t) = n[y(r,t)]$ where $y \equiv
r-r_0(t)$ and $r_0(t) = -ct$ for some constant speed $c$, we obtain
the ODE
\begin{equation}
D n^{\prime\prime}(y) -[(c+\nu) n(y)]^\prime + u(y)n(y)=0,
\label{EQ:StationSol}
\end{equation}
where $u(y)\equiv \left({\phi}^{-1}\Theta(-y)  -1\right)/\tau$. A
physically allowed solution of Eq.~(\ref{EQ:StationSol}) exists for
every $c \ge c_0 -\nu$, where $$c_0 \equiv
\sqrt{4D({\phi}^{-1}-1)/\tau}.$$ In fact, the smallest possible speed
$c_{\rm min}\equiv c_0-\nu$ is {\em selected} by the dynamics given a
reasonably compact initial distribution. Here, velocity selection
follows the familiar marginal stability
mechanism~\cite{REF:Collet1990}: The selected solution $n^*(y)$
(with the velocity $c_{\rm min}$)  is the one that decays
most sharply at the pulse front (i.e., the $r<r_0$ end) among all the
allowed solutions. Thus, as the front of the distribution
broadens from the initial condition, 
it first reaches the asymptotic decay of $n^*$. From then
on, the distribution stops broadening and moves with the speed $c_{\rm
min}$. Standard arguments show that this is equivalent to the
condition that the front be marginally stable in the frame of
reference moving with $c_{\rm min}$. Note that as $c_0 \propto
\sqrt{D} \propto \sqrt{\nu}$,  $c_{\rm min} <0$
when the mutation rate $\nu$ is
sufficiently large, indicating the worsening of the
overall affinity of the sequences due to accumulation of deleterious
mutations despite the presence of selection. These results apply also
to the more general Fermi function, as it is only the asymptotic
behavior of the growth term [i.e., $ U(r \ll r_0)$] that governs
velocity selection.


In evolutionary dynamics, the population size $N$ often plays a very
important role~\cite{REF:Peliti1997,REF:Drossel2001}. To see how $N$
enters,
we note that the mean-field equation (\ref{EQ:MeanField1}) has an
inconsistency in that at the very front of the moving pulse,
arbitrarily small $n$ gets the benefit of exponential amplification.
But in reality, the number of individuals is discrete so that $n$
should always be greater than $1/N$. To deal with this problem, a
cutoff procedure was proposed within the mean-field
framework~\cite{REF:Levine1996,REF:Derrida1997}.
Here we employ this procedure to estimate the effect of a finite
population on the evolutionary velocity~\cite{FNOTE:phasespace}.
Specifically, we modify the selection/amplification term $u(y)$ in
Eq.~(\ref{EQ:StationSol}) to $u(y)\Theta(n-N^{-1})$ (for $y<0$).  A
direct extension of the approach in Ref.~\cite{REF:Derrida1997} leads
to ${\delta c_0}/{c_0} \sim \frac {\pi^2}{2}/\ln^2 N$ for the
fractional change in $c_0$, which has the same scaling form as that
for the Fisher equation~\cite{REF:Derrida1997}. To test this result,
we ran simulations (with a modified mutational scheme to achieve a
constant drift) to measure the propagating speed of the pulse for
different population sizes $N$. 
Our finding of $\delta c/c \approx 0.06$
between $N = 5\times 10^3$ and $ 5\times 10^8$ is  in line with the
expectation and indicates that under typical experimental conditions,
the fluctuation effect due to finite population is insignificant.

We next examine the more realistic situation of finite sequence
length $L$. The important new effect is due to the $r$-dependence
in the  drift velocity $v(r)$ [see Eq.~(\ref{EQ:Dandv})], which, 
as the population approaches towards $r=0$, increasingly hinders its
advance. This can
already be appreciated  if we assume a
quasi steady-state dynamics and replace $\nu$ in the formula
for $c_{\rm min}$ with $v(r) = \nu - \gamma r$
[where $\gamma \equiv \frac43\nu_0$ according to (\ref{EQ:Dandv})]:
We find a {\em stable} stationary position, ${\overline r}^{\rm EQ}
\approx (\nu-c_0)/\gamma$ where $c_{\rm min} = 0$. Here we
identify this position naturally with the mean of the population
${\overline r}^{\rm EQ}\equiv \int rn ^{\rm EQ}(r) dr$.

To proceed with a more rigorous analysis,
we neglect the $r$ dependence in $D$ which has
only a small quantitative effect. Also, we assume that the
equilibrium position ${\overline r}^{\rm EQ}\gg 1$
so that  the boundary condition at $r=0$ can be
safely ignored.
\begin{figure}[b]
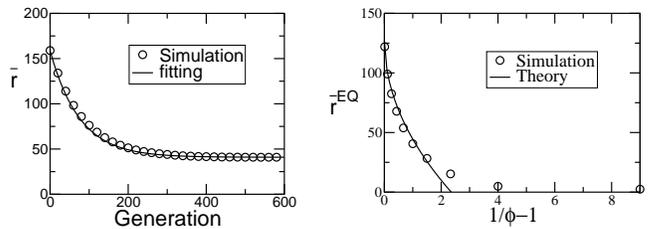

\includegraphics[height=1.15in]{MeanMismatchVTimePrune.eps}
\,\,\,\,
\includegraphics[height=1.15in]{EquilibriumPositionNew.eps}

\caption{\protect\flushing \label{MeanMismatchVTimePrune}(a)
Evolution of the mean mismatch ${\overline r}(t)$ for
$\phi=0.5$. The equilibrium distribution is used as the initial condition
to mitigate transient effects. The solid line is a single-exponential
fit using the theoretical value of $\gamma = \frac43\nu_0=.0133$.
(b) Equilibrium positions as a function of selection pressure
$\phi^{-1}-1$.  The line is the theoretical estimate
${\overline r}^{\rm EQ} = (\nu-c_0)/\gamma$,
using a generation time $\tau=2.77$ (obtained
by calibrating $c_0$ from theory with that from simulation).
}
\end{figure}
Returning to the mean-field equation (\ref{EQ:MeanField1}),
we use the same moving-pulse Ansatz as before except that we no longer
fix a linear time dependence to the threshold $r_0 (t)$.
This Ansatz produces a linear ODE for $r_0(t)$:
\begin{equation}
\gamma r_0(t) + {\dot r}_0(t) = \gamma r_0^{\rm EQ}
\label{EQ:AproachEqm}
\end{equation}
where $r_0^{\rm EQ}$ is the equilibrium threshold, so that a static
distribution can be achieved in the moving frame. The population mean
${\overline r}$ follows exactly the same motion (in fact, ${\overline
r}(t) \approx r_0(t)$ except when selection is very weak), i.e., a
single exponential with time constant $\gamma$ (which depends only on
the point mutation rate $\nu_0$). This is a generic result
independent of the details of the fitness function, as long as a
pulse solution exists for Eq.~(\ref{EQ:MeanField1}). The decay
constant $\gamma$ obtained from simulation of the discrete model is
in quantitative agreement with the expectation ($\frac43\nu_0$) for
weak selection ($1> \phi \gtrsim 0.25$); an example is shown in
Fig.~\ref{MeanMismatchVTimePrune}(a). In fact, the same qualitative
result (i.e., the existence of a shape-preserving pulse) holds for
strong selection as shown already in Fig.~\ref{shapeevolution} where
$\phi=0.1$.

The shape of the pulse, i.e. the equilibrium distribution $n^{\rm
EQ}$, is governed by the same ODE as Eq.~(\ref{EQ:StationSol}),
except that the constant velocity $c$ is replaced by
$-\gamma(y+r_0^{\rm EQ})$. The resulting equation again has a
continuum of physically allowed solutions, each having a different
shape and corresponding to a different equilibrium position $r_0^{\rm
EQ}$ (hence different ${\overline r}^{\rm EQ}$). Here we have an
interesting generalization of velocity selection to the selection
instead from a continuum of decelerating pulses. Again, starting from
a compact initial distribution, the dynamics selects the
solution~\cite{REF:Solutions} whose front ($y<0$) decays most
rapidly, (in this case a Gaussian falloff), whereas the other
solutions have a power-law front. 

The ${\overline r}^{\rm EQ}$
extracted from the selected solution agrees well with its heuristic
approximation of $({\nu}-c_0)/{\gamma}$ when ${\gamma}\ll 1$; see
Fig.~\ref{MeanMismatchVTimePrune}(b).
The theory is
quantitatively accurate~\cite{FNOTE:Shape} when the selection is not
too strong (e.g., $\phi^{-1} < 2.5$).  For very strong selection, the
equilibrium threshold position $r_0^{\rm EQ}$ approaches $r=0$ and the
boundary condition there needs to be taken into account.
When $c_0$ and $\gamma$ are expressed in terms of original parameters,
 ${\overline r}^{\rm EQ}=({\nu}-c_0)/{\gamma}$
suggests that for a population with sequence length $L\gg 1$,
the population pulse could stall at ${\overline r}^{\rm EQ} \gg1$.
In order for the population to reach the optimal at $r\approx 0$,
we need to increase the selection strength
(i.e., lowering $\phi$) or decrease the mutation rate so that
$\phi^{-1} \gtrsim 1 + \nu_0\tau L/2$,
to overcome the bigger entropic barrier associated
with longer sequences.

As there have been extensive studies of evolution on various
landscapes in the context of population
genetics~\cite{REF:Burger2000, REF:Peliti1997, REF:Drossel2001}, it
is worth comparing the dynamical behavior of competitive evolution
with that of more common evolutionary models. The traditional study
of evolution focuses on fixed fitness landscapes, where every
genotype (e.g., sequence $\vS$) has a predetermined absolute fitness
value (i.e., the reproductive rate of the sequence $\vS$).
Competitive evolution is different in that it is subject to a
{\em dynamic} fitness landscape. That is, the fitness is measured
relative to a dynamic selection threshold and progress towards the
best binding sequence occurs via competition among the currently
existing genotypes --- being better is all-important, not being best.
This aspect of competitive evolution leads to qualitatively different
dynamical behavior. For comparison, we can consider the simplest and
most widely studied fixed landscape, i.e., the smooth \lq\lq Mt.\
Fuji\rq\rq\ landscape~\cite{REF:Burger2000, REF:Peliti1997,
REF:Drossel2001, REF:Woodcock1996}, where each nucleotide contributes
independently and additively to the {\it fitness} of the sequence,
thus forming a landscape on which fitness rises steadily toward a
single peak. For infinite sequence length,
the mean-field theory fails in that it produces an unphysical,
run-away solution~\cite{REF:Levine1996} due to the unlimited
growth rate of $n$ at the high fitness states
~\cite{REF:Woodcock1996,REF:Levine1996,REF:Kessler1997andRidgeway1998}, 
and a finite population has
a traveling speed that is essentially proportional to population
size~\cite{REF:Kessler1997andRidgeway1998}. For finite
sequence length, the finite population dynamics is
orders-of-magnitude slower in reaching equilibrium than the (now
non-divergent) mean-field
prediction~\cite{REF:Kessler1997andRidgeway1998}. In contrast, 
finite population effects merely cause a small correction
for the competitive evolutionary process.

To summarize, we investigated the dynamics of competitive evolution
in the context of molecular evolution experiments. The major result
concerns the existence and properties of a shape-invariant population
pulse which propagates towards an eventual equilibrium configuration.
Analytical results on the motion of the pulse obtained from the
mean-field equation are in good agreement with
simulations.  Also, corrections due to finite population size are
shown to be insignificant.  An interesting aspect of our findings is
the convergence of the evolution process to a solution far from
optimal (i.e., ${\overline r}^{\rm EQ} \gg 1$), if the selection
strength is not sufficiently strong or mutation rate not sufficiently
low.  In general, competitive evolution is rather different from the
usual picture of climbing a fixed fitness landscape.  This approach
may be applicable more generally, e.g. to natural evolution in cases
where competition for scarce resources is the primary driving force,
as an organism only needs to be more efficient than its competitors
to win the battle for evolutionary survival.

We thank M.\ Kloster for detail comments and discussions.  We also
acknowledge helpful discussions with D.\ Kessler, Q.\ Ouyang, L.\ Tang
and J.\ Widom. This research is supported by the NSF through Grants
 DMR-0211308 and MCB-0083704.

%

%

\begin{thebibliography}{99}
%
\bibitem{REF:Arnold2001}
E. T. Farinas, T. Bulter and F. H. Arnold, Curr. Opin. Biotechnol.
\textbf{12}, 545 (2001).

\bibitem{REF:Stemmer1994}
W. P. C. Stemmer, Nature \textbf{370}, 389 (1994).

\bibitem{REF:Collet1990}
See, e.g., P. Collet and J. -P. Eckmann, \textit{Instabilities and Fronts
in Extended Systems} (Princeton University Press, Princeton, 1990).

\bibitem{REF:Dubertret2001}
B. Dubertret, S. Liu, Q. Ouyang and A. Libchaber,
Phys. Rev. Lett. \textbf{86}, 6022 (2001).

\bibitem{REF:Lodish2000}
See, e.g., H. Lodish {\it et. al.}, \textit{Molecular cell biology}, $4$th ed
(W.H. Freeman \& Co., New York, c2000).

\bibitem{REF:Widom1999}
A. Thastrom {\it et al.},
J.\ Mol.\ Biol.\ \textbf{288}, 213 (1999).

%
\bibitem{REF:VonHippel}
P. H. von Hippel and O. G. Berg,
Proc. Natl. Acad. Sci. \textbf{83},  1608 (1986).


\bibitem{fields}
D.S.\ Fields, Y.\ He, A.Y.\ Al-Uzri and G.D.\ Stormo,
J.\ Mol.\ Biol.\ \textbf{271}, 178 (1997).

\bibitem{gmh}
U. Gerland, J.D. Moroz and T. Hwa, Proc.\ Natl.\ Acad.\ Sci.\
\textbf{99}, 12105  (2002).

\bibitem{nonspecific} To focus on the steady state dynamics,
we have ignored
``nonspecific'' protein-DNA binding~\cite{REF:VonHippel,gmh}.
This would affect the
initial stage of evolution which plays a dominant role
in the experiment of Ref.~\cite{REF:Dubertret2001}.

\bibitem{REF:Gerland2001}
U. Gerland and T. Hwa, J. Mol. Evol. \textbf{55}, 386 (2002).

\bibitem{REF:Peliti1997}
L. Peliti, cond-mat/$9712027$.
\bibitem{REF:Drossel2001}
B. Drossel, Adv. Phys. \textbf{50}, 209 (2001).




\bibitem{REF:Kloster2}
M. Kloster and C. Tang, manuscript in preparation.

\bibitem{REF:Levine1996}
L. S. Tsimring, H. Levine and D. A. Kessler, Phys. Rev. Lett. \textbf{76}, 4440 (1996).

%
\bibitem{REF:Derrida1997}
E. Brunet and B. Derrida, Phys. Rev. E \textbf{56}, 2597 (1997).


\bibitem{REF:Kessler1997andRidgeway1998}
D. A. Kessler, H. Levine, D. Ridgway and L. Tsimring,
J. Stats. Phys. \textbf{87}, 519 (1997); {\it ibid}., \textbf{87}, 519 (1997).
%

\bibitem{FNOTE:phasespace}
Strictly speaking, $1/N$ is the cutoff value in the sequence
space ($\vS$) rather than the mismatch space ($r$).
As there is only selection in the mismatch space, the distribution
forms a compact packet and undergoes random walk in the orthogonal
direction. Hence, we expect the cutoff $1/N$ to be valid
up to a constant even in mismatch space.

\bibitem{REF:Solutions}
The selected solution can be expressed via the parabolic cylinder
function
${\mathcal D}_{p}(\tilde y)$, with $ n^{\rm EQ} \propto {\mathcal
D}_{p_{+}}\left( {\tilde y} \right) e^{- {\tilde y}^2 /{4}}$
[${\mathcal D}_{p_{-}} \left({-\tilde y}\right) e^{- {{\tilde y}^2
}/{4}}$] for $y >0$ ($y\le 0$), where $p_{+}=-1/\gamma\tau$ [$p_{-} =
(\phi^{-1}-1)/\gamma\tau$] and ${\tilde y} \equiv (y+r_0^{\rm
EQ}-\nu_0/\gamma) \sqrt{\gamma/D}$. $r_0^{\rm EQ}$ is determined by
matching conditions at $y=0$.


\bibitem{FNOTE:Shape} 
Although the analytical solution gives a fair
description of the motion of the pulse, it offers a pulse shape much
broader than observed in simulation. A better description of the shape
is presented in~\cite{REF:Kloster2}.

\bibitem{REF:Burger2000}
See, e.g, R. B\"urger, \textit{The Mathematical Theory of Selection,
Recombination and Mutation} (John Wiley \& Sons, Chichester, UK, 2000).


\bibitem{REF:Woodcock1996}
G. Woodcock and P. G. Higgs, J. Theor. Biol. \textbf{179}, 61 (1996).





\end{thebibliography}
\end{document}